\documentclass[journal]{IEEEtran}

\usepackage{enumerate}
\usepackage{paralist}
\usepackage{graphicx}
\graphicspath{{Figures/}}
\usepackage{subfigure}
\usepackage{cite}
\usepackage{array}
\usepackage{booktabs}
\usepackage{comment}

\usepackage{amsmath}
\usepackage{mathtools}
\usepackage{enumerate}
\usepackage{paralist}
\usepackage{array}
\usepackage{booktabs}
\usepackage{comment}
\usepackage{cases}
\usepackage{algorithm}
\usepackage{algorithmic}
\usepackage{mathrsfs} 
\usepackage{url}
\usepackage{amssymb}
\usepackage{amsthm} 

\usepackage{bigints} 

\usepackage{epstopdf} 

\usepackage{color} 

\begin{document}

\title{Social Behavior in Bacterial Nanonetworks: Challenges and Opportunities}

\author{Monowar Hasan, Ekram Hossain, Sasitharan Balasubramaniam, and Yevgeni Koucheryavy

} 
\date{}

\maketitle

\begin{abstract}

Molecular communication promises to enable communication between nanomachines with a view to increasing their functionalities and open up new possible applications. Due to some of the biological properties, bacteria have been proposed as a possible information carrier for molecular communication, and the corresponding communication networks are known as \textit{bacterial nanonetworks}. The biological properties include the ability for bacteria to mobilize between locations and carry the information encoded in Deoxyribonucleic Acid (DNA) molecules. However, similar to most organisms, bacteria have complex social properties that govern their colony. These social characteristics enable the bacteria to evolve through various fluctuating environmental conditions by utilizing cooperative and non-cooperative behaviors. This article provides an overview of the different types of cooperative and non-cooperative social behavior of bacteria. The challenges (due to non-cooperation) and the opportunities (due to cooperation) these behaviors can bring to the reliability of communication in bacterial nanonetworks are also discussed. Finally, simulation results on the impact of bacterial cooperative social behavior on the end-to-end reliability of a single-link bacterial nanonetwork are presented. The article concludes with highlighting the potential future research opportunities in this emerging field.

\end{abstract}


\section{Introduction} \label{sec:intro}

Molecular communication is a paradigm that aims to develop communication systems at the nanoscale \cite{AkyildizComNet}. In order to ensure efficiency and biocompatibility, the objective of this new communication paradigm is to develop  communication systems by utilizing components that are found in the nature. Such a communication system will include at least one transmitter nanomachine\footnote{A nanomachine is a miniature device consisting of a set of molecules which are able to perform very simple computational tasks. A nanomachine could consist of several integrated components, such as the control unit, transmitter and receiver, power unit, sensing unit, etc. For detailed overview of a nanomachine, we refer the reader to \cite{AkyildizComNet}.} which encodes information into molecules (i.e., ions, Deoxyribonucleic Acid [DNA] molecules). These molecules will be transported to the receiver and decoded. Example models of molecular communication that have been proposed include molecular diffusion of information molecules \cite{physical_ian} and those using active carriers such as bacteria  \cite{ian_bbcn, sasi_bbc}. Enabling communication at the nanoscale and interconnecting the molecular nanonetworks to the Internet could provide opportunities for a new generation of \textit{smart city} and \textit{health-care} applications. Examples of these applications include:

\begin{itemize}

\item \textit{Environmental sensing}: The future smart city envisions more accurate and efficient sensing techniques for the environment. This sensing process may include early detection of pathogens that may affect the crops or live stocks. Since bacteria are found widespread within the environment, they can serve as information carriers between nano sensors, and collect information at fine granular scale. 

\item \textit{Biofuel quality monitoring}: One alternative source of energy is the conversion of biomass to fuel production. Recently, scientists have experimentally shown how engineered bacteria could turn glucose into hydrocarbon that are structurally identical to commercial fuel \cite{bio_fuel}. Therefore, utilizing bacterial nanonetworks could improve the quality of biofuel production, and at the same time provide accurate quality control.

\item \textit{Personalized health-care}: The process of early disease detection within the human body is a major challenge. Detecting diseases at an early stage can provide opportunities of curing the condition and prevent further spreading of the disease. Since bacteria are found in the gut flora, embedding nanonetworks into the intestine can provide fine granular sensing at the molecular scale. Besides sensing, the bacterial nanonetworks can also provide new methods for targeted drug delivery.

\end{itemize}

In this article, we focus on the use of bacteria to transport DNA-encoded information between the nanomachines. In a bacterial nanonetwork, bacteria are kept inside the nanomachines and then released to commence the information transmission process \cite{sasi_bbc}. While numerous works have investigated the feasibility of bacterial nanonetworks (e.g., \cite{ian_bbcn, sasi_bbc}), the communication models used in the earlier works have not considered bacterial social behavior. Bacteria usually co-exist as a community, which at times could consist of multi-cellular community. The community structure enables bacteria to cooperate and co-exist in varying environmental conditions. However, extreme environmental conditions (e.g., scarce resources) could also lead to competitive and non-cooperative behavior among the bacteria species. This usually results in each species developing strategies for survival. Since bacterial nanonetworks will rely on bacteria carrying messages between the different nanomachines, the social properties can affect the performance and reliability of the bacterial nanonetworks. 

We provide an overview of various bacterial social behavior and the challenges as well as opportunities they create in the context of the reliability of communication in bacterial nanonetworks. An analogy can be drawn between the social-based bacterial  nanonetworks and the social-based Delay-Tolerant Networks (DTNs), where the social behavior of people can affect the performance of mobile ad hoc networks. The key contributions of this article can be summarized as follows:

\begin{itemize}
\item We review and analyze the impact of bacterial social behaviors on the performance of the nanonetworks. We describe the various challenges and opportunities that arise due to the bacterial social behavior in such networks. 

\item Using computer simulations, we demonstrate the use of bacterial cooperative social behavior that help to entice the bacterial motility towards the destination. The results from the simulations show that the cooperation can substantially improve the network performance.

\item This article creates a new direction of research to address the challenges in future molecular nanonetworks that utilize bacteria as information carrier. In particular, the article provides a guideline to exploit bacterial social properties in a dynamic environment to improve the communication performance.
\end{itemize}


The remainder of the article is organized as follows: after an overview of the bacterial nanonetworks, we provide an introduction to the communication mechanisms among different bacterial species. This is followed by a review of the social properties of bacterial community. Then, we describe the challenges and opportunities that arise due to the bacterial social behavior from the perspective of the communication performance in bacterial nanonetworks. We present results from simulation studies to evaluate the effect of  dynamic social behavior on the performance of the communication nanonetwork. To this end, we highlight several future research scopes in this emerging multi-disciplinary field before we conclude the article.

\section{Bacterial Communication Processes and Nanonetworks} \label{sec:com_process}

\subsection{Bacterial Communication Processes}

The  social behaviors of bacteria result from their communication capabilities. Again, this communication results from bacterial linguistics, which is enabled by emitting various biochemical agents\footnote{For details of the biochemical signaling agents and bacterial linguistic communication, refer to \cite{bacteria_lingusitic} and references therein.}. The communication process of the bacteria is not only limited to bacteria of the same species, but it can also extend to multi-colony and inter-species communication \cite{multicolony-1}. 
Recent studies have identified that inter-species message-passing occurs quite regularly in multi-species \textit{biofilms}. The biofilms refer to surface-attached densely-populated communities formed by the bacteria. For instance, larger population of antibiotic resistant cells within a bacterial population can emit chemical signals (e.g., small molecules) to increase antibiotic resistance in less resistant cells. These small molecules are not limited to protect the cells within the same species, but can also extend to other species \cite{antibiotic_uwo}. 

\begin{table*}[t]
  \centering
  \caption{Communication types in bacterial nanonetworks}
  \label{tab:comm_prop}
  \begin{tabular}{p{2.5cm} p{5.8cm} p{7.9cm}}
  \hline 
Communication type & Process & Example of bacteria used \\ \hline \hline
1. Molecule-based & \\
&  Local diffusion of molecules & \emph{Streptoccocus pneumonia} (communicates through short peptides that contains chemical modifications) \\
& &\emph{Vibrio} (produces \emph{Acyl-Homoserine Lactone [AHL]} which leads to quorum sensing) \\
& & \emph{Streptomycetes} (communicates using structural analog of AHL)\\ \hline
2. Plasmid-based &  &\\
	\hspace{1.5em}  & Conjugation (exchanges plasmids through physical pilus connection) & \emph{Escherichia coli} (\emph{E. coli})  \\
	\hspace{1.5em} & Nanotubes (transfers ions and plasmids) &  \emph{E. coli} and \emph{Bacillus subtilis}\\
	\hspace{1.5em} & Bacteriophage (uses virus to transfer plasmids) &  \emph{E. coli} \\ \hline
	    
  \end{tabular}
  
\end{table*}

Table \ref{tab:comm_prop} presents examples of communication mechanisms for different bacterial species. There are mainly two different mechanisms of communication: (i) molecule-based communication and (ii) DNA-based communication, which are briefly described in the following. 

\begin{enumerate} [\itshape(i)]

\item \textit{Molecule-based communication}: In molecule-based communication, bacteria emit molecules from their membrane, where these molecules would diffuse randomly within the local environment. Based on their random diffusion, the molecules will be eventually  picked up by the bacterial population that are in close vicinity. The communication process is executed primarily by chemical signals using the following entities:  \textit{signaling cell}, \textit{target cell}, \textit{signaling  molecule}, and \textit{receiver protein} \cite{book:molecularbiology}. The signaling cell is responsible for diffusing the molecules to one or more target cells, where the information is encoded into the signaling molecule. The receiver protein is responsible for decoding and transferring the message to the inter-cellular plasma. As shown in Table \ref{tab:comm_prop}, there are various types of chemical molecules that are used for molecule-based communication.

\item \textit{Plasmid-based communication}: The \emph{plasmids} are genetic molecules that are carried by the bacteria. In the plasmid-based communication, the bacteria are able to transfer plasmids between each other. Table \ref{tab:comm_prop} presents example mechanisms that are used for this type of communication. The \emph{conjugation} process involves bacteria forming a physical connection using the \emph{pilus}. The pili (plural of pilus) are tubular proteins that stem from the bacteria membrane. When the bacteria are in close proximity, the pilus will form a physical connection which facilitates the transfer of plasmid from one bacterium to another. The nanotubes are also physical connections formed between the bacteria on a solid surface \cite{mediate_bacterial}. The contents that are passed between the bacteria using nanotubes are not only limited to non-conjugative plasmids, but they also include ions and proteins. Another approach for transferring genetic content is through \emph{Bacteriophages}. The bacteriophages are a kind of virus that can be formed within the bacteria. In \cite{ortiz2012engineered}, experimental tests show that \textit{E. coli} are able to transfer information using bacteriophages. The bacteria will emit the phages with encapsulated plasmids, where the phages will diffuse within the medium and be picked up by the bacteria swimming within the vicinity. Unlike conjugation, the probability of plasmid transfer using this approach will lead to a higher number of bacteria receiving the plasmid since physical connection and contacts between the bacteria are not required. 

\end{enumerate}

\subsection{Bacterial Nanonetworks}

\begin{figure}[!t]
	\centering
		\includegraphics[width=2.5in]{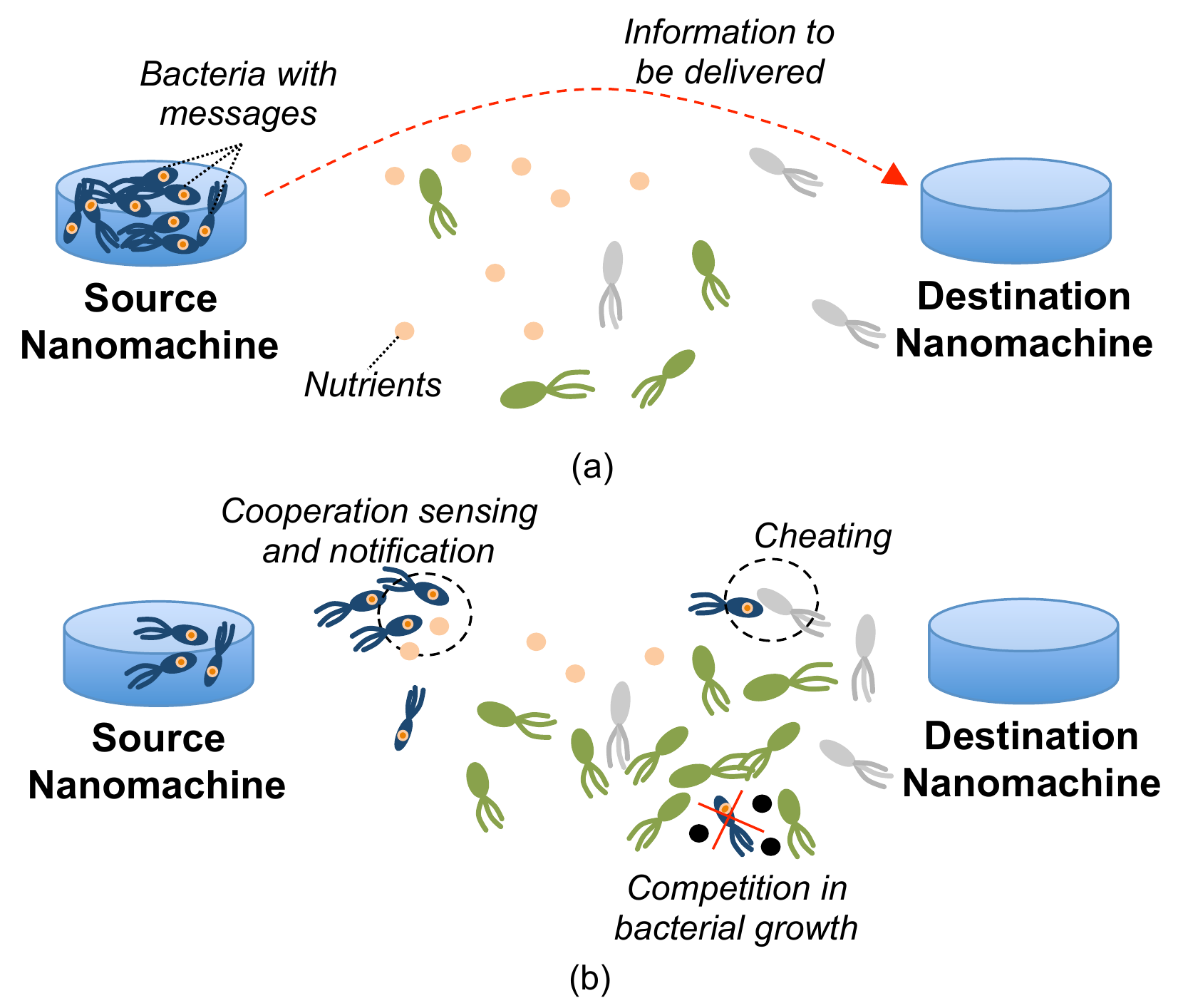}
	\caption{Social behavior in a bacterial nanonetwork: (a) Overview of communication process, (b) Influence of social behaviors during information transmission.}
	\label{fig:BacMolNets}
\end{figure}

Bacteria have a number of interesting properties that make them ideal as information carriers in nanonetworks. Besides the communication processes described above, the bacteria also have the ability to mobilize themselves through a medium (e.g., swimming, gliding, twitching). This motility provides an added benefit over approaches that utilize pure diffusion to propagate information molecules in molecular nanonetworks \cite{{physical_ian}}. By incorporating these properties, Fig.~\ref{fig:BacMolNets} presents the vision of a bacterial nanonetwork. The communication process involves the bacteria encoded\footnote{A brief overview of DNA-encoding of information into the bacterium is provided in \cite{ian_bbcn} and \cite{AkyildizNanoMotor}.} with information plasmid released from a source nanomachine. The bacteria will mobilize themselves towards the destination nanomachine and unload the information plasmid.

The process of encoding of information for the bacterial nanonetwork could be based on the technique proposed in \cite{AkyildizNanoMotor}. The first step is encoding of digital information into a genetic sequence. The DNA molecule is composed of two polymers of nucleotides. Each nucleotide contains one of four possible bases: Adenine ($\mathtt{A}$), Cytosine ($\mathtt{C}$), Guanine ($\mathtt{G}$), or Thymine ($\mathtt{T}$). Therefore, the encoding process could be as simple as setting $\mathtt{G}$ and $\mathtt{T}$ bases to one with $\mathtt{C}$ and $\mathtt{A}$ bases equal to zero, i.e., $\mathtt{GT} = 1$, $\mathtt{AC} = 0$. Hence, if the data we want to transmit is $10110011$, the corresponding genetic code will be $\mathtt{GATTACTG}$. Once the information is encoded into a DNA molecule, this could then be inserted into a plasmid, where the insertion process could be through \emph{ligation}. The plasmid could then be inserted into the bacteria through the process of \emph{transformation}. Since the process of encoding is out of the scope of this article, we only consider that the information is pre-encoded into the bacteria and is placed into a source nanomachine. When the bacteria reaches the destination nanomachine, the plasmid could be offloaded through the process of \emph{conjugation}. The process of conjugation involves creating a physical connection that allows the plasmid to be copied.

This particular form of molecular communication is most ideal for transferring DNA-encoded information between the nanomachines. However, as reflected in Fig. \ref{fig:BacMolNets}, dynamic social behaviors (e.g., cooperation, competition etc.) of the bacteria may influence reliability of information transfer in the nanonetwork. In the following section, we will discuss various social properties of bacteria, and how they can influence the performance of the nanonetwork. 

\section{Social Behavior of Bacteria} 
\label{sec:social}

Although bacteria can be utilized to transfer the information between the nanomachines, the challenge is to ensure that they can successfully reach the destination nanomachine. In addition to the stochastic mobility nature of the bacteria, the other challenge for a reliable bacterial nanonetwork arises due to social interaction among bacteria within the environment. These social interactions could either support or act as a hindrance for information transfer. Recent studies of bacterial species in biofilms have revealed diverse complex social behaviors, including cooperative and non-cooperative social behaviors.  Bacteria are also able to store information, perform decision making and learn from past experiences collectively. In this section, we review some of the social behaviors of bacteria that demonstrate a complex and coordinated social life. We subdivide the social behaviors as \textit{cooperative} and \textit{non-cooperative} behaviors which are summarized in Table \ref{tab:bact_behave}.


\subsection{Cooperative Behaviors} \label{subsec:coop}

Bacteria cooperate to protect themselves from enemies, secure nutrients, enable reproduction or mobilization to new favorable locations. The bacterial cooperation could be through an egalitarian process. In this process, all individuals contribute and gain more or less equally, or it involves \textit{division of labor}, i.e., individuals engaging in different tasks might obtain different rewards \cite{evo_micro, physics2complex, bacteria_lingusitic}. From laboratory experiments, it has been observed that the bacterial colony performs collective sensing, distributed information processing, division of labor, and support gene-regulation of individual bacteria in the colony \cite{physics2complex}. The examples of bacterial cooperative behaviors can be represented (but not limited to) by the following phenomena:


\begin{enumerate} [\itshape(i)]

\item \textit{Cooperative hierarchical organization:}  As an adaptive response to environmental stress, bacteria can form complex spatial organization of the colony by utilizing pattern formation mechanisms (e.g., see Figs. 1-3 in \cite{physics2complex}). Some bacterial strains organize their colonies by generating modules in response to nutrient-depletion and hard substrates. An example of cooperation is the production of lubricating layer of fluids during adverse conditions. By adjusting the lubricant viscosity, bacteria can keep the population density high for protection as well as efficient use of resources. Within the colony, the bacteria can form branching patterns through combined action of different chemotactic strategies. Utilizing branching patterns, individual bacterium can make maximal use of resources, and interestingly, the decisions are made cooperatively \cite{physics2complex}.

\item \textit{Cooperative sensing and notification:} Bacteria are able to emit notification signals when they detect varying levels of nutrients. Bacteria communicate about levels of nutrients by means of attractive and repulsive \textit{chemotactic signaling}\footnote{Chemotactic signaling is a chemotaxis response to chemicals produced by the bacteria.}. In attractive signaling, bacteria emit food-like molecules to entice other bacteria to move towards them. Using repulsive chemotactic signaling, bacteria emit chemicals which drive the overall colonial growth away from themselves. As a consequence, other colonial members can stay away from regions of low nutrients or harmful chemical imbalances \cite{jacob_cognition}. Another example is \textit{Quorum Sensing} (\emph{QS}) \cite{qs-1} which is a collaborative communication process that uses advanced bacterial sensing capabilities. The benefit of QS is twofold: it helps bacteria to coordinate processes (e.g., formation of biofilms) that would be inefficient in single cells; and also enables the bacteria to sense the density of population.
An example of QS-based bacterial communication is apparent in \textit{Vibrio fischeri}\footnote{\textit{Vibrio fischeri} is a gram-negative, rod-shaped bacterium found in the marine environments.}. The bio-luminescent element produced by \textit{V. fischeri} would not be visible if it were produced by a single cell. However, by utilizing QS, the bio-luminescent is produced only when the population density exceeds a threshold \cite{physics2complex}.

\item \textit{Foraging:} A form of cooperation found in bacteria
involves food acquisition. For example, \emph{Myxobacteria}\footnote{The Myxobacteria are a group of soil bacteria.} collectively develop mass attacks on microbial prey in order to consume them as food source. Another form of apparent cooperation in foraging includes complex growth forms in some bacterial colony structures that maximize their feeding ability \cite{evo_micro}. The foraging process could integrate sensing as well.  One example of bacterial collective sensing and division of labor is described as follows: some of the members (e.g., foraging bacteria) in the colony advance to the edge of the colony and upon chemical sensing of food (e.g., nutrient) source, the information is returned to the colony. The bacterial colony then collectively expands by cell division and mobilize through gliding to the newly detected food source \cite{bacteria_multicellular}.

\item \textit{Protection and immolate:} Bacteria are also able to collaborate in order to protect the best interest of the population. During nutrient scarcity, bacteria at the edge of the colony can coordinate the limited reproduction process in order to consume less nutrients. In other cases, small proportion of the cells in the colony can suicidally produce large quantities of chemicals (e.g., colicins), which kill bacteria of the competing species in order to save their own population \cite{evo_micro}. 

\item \textit{Collective memory, learning and information processing:} Bacteria can generate an erasable, collective inheritable memory that they have learnt from past experiences \cite{physics2complex, jacob_learning, jacob_cognition}. For example, upon encountering antibiotic stress, bacteria employ a particular strategy that reshapes the colony pattern\cite{physics2complex}. This strategy will lead to enhanced cooperation among the bacteria by intensifying the chemotactic attraction that leads to large forms of vortices. The vortices are branching patterns that bacteria create through cooperation. By creating the large stems of vortices, the antibiotic within the environment is diluted through the lubricating fluid excreted by the bacteria \cite{jacob_cognition}.
 
A bacteria colony is also able to conduct distributed information processing, where each bacterium is capable of storing, processing, and interpreting information \cite{jacob_learning, jacob_cognition}. When coping with environmental changes, the bacteria sense the environment and perform internal information processing as well as coordinate the information by means of biochemical communication (e.g., QS). Learning and 
information processing in a colony happen at two levels of 
abstraction. On the first level, the biochemical exchanges occur among individuals as well as between the colony and the environment that generates and accesses the collective memory of the colony. The second level is formed by the intra-cellular communication network that analyzes and interprets the information extracted from the environment.

\end{enumerate}



\subsection{Non-cooperation and Clashes in Bacterial Strategy} 

Cooperative behavior is beneficial for the individuals as well as the general population by providing public goods. Public goods refer to the secreted products that are costly to synthesize but benefit other cells in the population. For example, biofilms formed by the bacteria comprise not only of bacterial cells but also various compounds that the cells release into the surrounding environment. Many of these compounds are diffusible substances (including digestive enzymes and chelating compounds) which aid in nutrient acquisition. However, there are situations where certain \textit{opportunistic individuals} (i.e., non-cooperators or cheaters) prefer not to cooperate in order to obtain the advantage of the group's cooperative effort. An overview of non-cooperative phenomena observed in bacterial population are as follows:


\begin{enumerate} [\itshape(i)]

\item \textit{Cheating:} There has been an increasing number of experiments (e.g., \cite{yeast_cheating2}) on non-cooperative behavior of the bacterial population. Study on bacterial non-cooperation (i.e., cheating) during QS has shown a reduction in the population growth and the density of the biofilm. This non-cooperation in turn forces the cooperator cells to increase their level of cooperation. In the case of the biofilm, once the non-cooperators start to outnumber the cooperative bacteria, this could lead to the total collapse of the entire biofilm structure. This situation leads to a reduction in the population's overall productivity \cite{QS_cheating}, and in the worst case can collapse the total population. Researchers comment that the non-cooperative behavior of bacteria can be modeled using game theory, such as the \textit{prisoner's dilemma} game, where cheating will be the most rewarding strategy independent of the opponent's choice. However, recent study in \cite{yeast_cheating1} shows that cheating can be profitable, but not necessarily the best strategy if other members are cheating as well. 
 
\item \textit{Competition in bacterial growth:} Bacteria can affect each others' growth by unfairly consuming the limited resources (e.g., growth nutrient). By dividing rapidly, the non-cooperative bacteria can obtain a larger share of such resources and reduce the nutrient availability for other
members of the population. The effect of this selfish consumption leads to overall inefficient resource utilization (i.e., less public goods produced per unit nutrient consumption). Consequently, fast growth may decrease the biofilm's total productivity due to a trade-off between growing quickly (i.e., benefiting non-cooperators) and growing efficiently (i.e., benefiting the whole colony). Examples of no-win competitions can be seen in several bacteria. For example, \textit{E. coli} are able to produce and secrete specific toxins (e.g., colicines) that inhibit the growth rate of other bacteria \cite{jacob_cognition}. The toxin producing strain kills the colicin-sensitive strain that outcompetes the colicin-resistant strain. 

\item \textit{Clashes of bacterial strategy:} Bacteria can use a variety of sophisticated and dynamic strategy when the collective behavior is challenged by non-cooperators. For example, they can single out non-cooperators by collective alteration of their own identity into a new gene expression state  
\cite{jacob_cognition}. This clash with non-cooperators is beneficial to the group since it helps the bacteria to improve social skills for better cooperation.

\end{enumerate}

\begin{table}[t]
  \centering
  \caption{Bacterial social behaviors}
  \label{tab:bact_behave}
  \begin{tabular}{p{3.5cm} p{4.4cm}}
  \hline 
Bacterial behavior & Probable reasons \\ \hline \hline
1. \emph{Cooperative} & \\
	\hspace{1.2em} Hierarchical organization & Defense from unfavorable conditions, adaptation to environmental changes; protection from antibiotics \\
	\hspace{1.2em} Foraging & Food acquisition, dynamic adaptation to changes in the nutrient levels \\
	\hspace{1.2em} Cooperative sensing & Adaptation to environmental changes, cooperation for collective food acquisition \\
	\hspace{1.2em} Protection & Resource (e.g., nutrient) limitations, defense from predators	\\
	\hspace{1.2em} Learning & Self adaptation, responses to the environmental stress, resistance against antibiotics 	\\ \hline
2. \emph{Non-cooperative} &  \\
    \hspace{1.2em} Cheating and strategy clash & Greedy utilization of public goods, scarcity of resources, improvised inter-species cooperation  \\
    \hspace{1.2em} Competition & Scarcity of public goods, nutrient depletion \\ \hline
  \end{tabular}
\end{table}

\section{Challenges and Opportunities in Bacterial Nanonetworks} \label{sec:challanges}

As described in the introduction, our  aim is to utilize bacteria as an information carrier between the nanomachines in order to enable molecular communication. However, the uncertain conditions as well as the non-cooperative social behavior could affect the bacteria carrying the message. On the other hand, the cooperative behavior could be beneficial for the performance of the nanonetworks. The cooperative behavior could lead to population survivability, which implies that this will support the bacteria carrying the message to successfully arrive at the destination nanomachine. As described earlier, an example of this is when the cooperation allows the bacteria to form fluidic boundaries in order to protect other bacteria in the population. A key issue, however, is the non-cooperative behavior of the bacteria which could affect the information transmission probability. In this situation, the bacteria released from the transmitter nanomachine, which are carrying the message, are vulnerable and  may not successfully arrive at the receiver nanomachine. 

In the following we list a number of challenges and opportunities arising due to the social behavior of bacteria that can affect the communication performance.

\subsection{Changes in the Quantity of Nutrients}
Bacteria, similar to most organisms, rely on environmental nutrients for survival. The previous section described how cooperative behavior between the bacteria can enable nutrients to be discovered (e.g., sensing) as well as fair delivery (e.g., foraging). However, we have also seen that depletion of nutrients can lead to the bacterial species switching towards negative behavior. This will not only affect multi-species bacteria, where one species may try to kill off another species, but also amongst the same species. In the context of molecular communication, the bacterial species that is killed maybe responsible for the information transfer. Therefore, the design of communication between the nanomachines will need to consider fluctuations of nutrients in the environment, and obtain solutions to cope with the bacteria that are trying to eliminate each other. 

One approach to mitigate this situation and turn this into an opportunity is to ensure a stable environment. Stable environment with sufficient nutrients minimizes the competition among the bacteria and hence improves communication reliability. The nanomachines that will release the bacteria with the embedded information could also encapsulate nutrients from the nanomachine. These nutrients can be released at the same time as the bacteria with the encoded information. Once the nutrients are diffused into the environment,  the bacteria with the encoded information can reproduce in numbers. This will enable the species of the bacteria carrying the messages to possibly outnumber the other competing species, in the event they decide to release toxins to kill the other species. 

\subsection{Changes in the Behavior Due to Cheating}
Although the changes in the quantity of nutrients can affect the environment, this is not the only factor that can change the social interactions of the bacteria. As has been described earlier, certain bacteria can switch to selfish behavior in order to seek individual benefit. The learning capabilities of the bacteria may also lead to the behavioral switching. For instance, if the bacteria are initially cooperating and sense a high enough density of population within the environment, they may decide to switch the behavior believing that their change may not be detected by the general population. In such a case, if a nanonetwork is embedded within a biofilm and this biofilm structure fails, this could lead to a full breakdown of the nanonetwork. 

One solution to mitigate  this problem is to ensure that the environment contains an optimum density of nanomachines forming the network so that the network will be robust  under failures. Therefore, in the event of biofilm breakdown, the nanonetwork may be subdivided into sub-networks.

\subsection{Destructive Communication}

Previous discussions have described the destructive effects of non-cooperative bacteria on the communication performance.
One method to improve the communication performance is to apply antibiotics within the environment to kill off bacteria that are harmful. However, the bacteria could develop resistance to the antibiotics and this resistance could be through a gene within a plasmid. Through the conjugation process these plasmids with  resistance to antibiotics could be passed between the bacteria. Note that the conjugation process is generally beneficial for bacterial nanonetworks since it increases the quantity of messages that could be delivered to the destination nanomachine. Since this could be negatively utilized by the harmful bacteria, both the positive and negative effects of conjugation should be taken into consideration when designing bacterial social networks.

In order to curb the non-cooperative behaviors, the nanomachines within the environment could also dispense antibiotics. This will require the bacteria carrying the plasmid with the encoded information to also possess the antibiotic resistance genes. In the event that the bacteria carrying legitimate plasmids are conjugated with the other bacteria, they will also transfer the plasmids with the antibiotic-resistant genes. When the antibiotics are diffused before any transmission, this will ensure that the non-cooperators without the resistance genes are eliminated from the environment. Therefore, this will lower the probability of conjugating with the non-cooperators. 

\section{Simulation of Cooperative Social Behavior in Bacterial Nanonetworks} \label{sec:sim}

In order to observe the impact of cooperative bacterial social interaction, in this section, we evaluate the communication performance in a bacterial nanonetwork through simulations.\footnote{By the term ``simulation'' we refer to the computer simulations based on the mathematical model and assumptions presented in the following subsection (Section \ref{subsec:sim_assmp}). We use MATLAB to develop the simulator.} We compare and analyze the results with the bacteria-based nanonetwork approaches that have been proposed in the existing literature (e.g., \cite{ian_bbcn, sasi_bbc}), where cooperation is not considered. 

\subsection{Communication Scenario} \label{subsec:sim_assmp}

\begin{figure}[!t]
	\centering
		\includegraphics[width=2.5in]{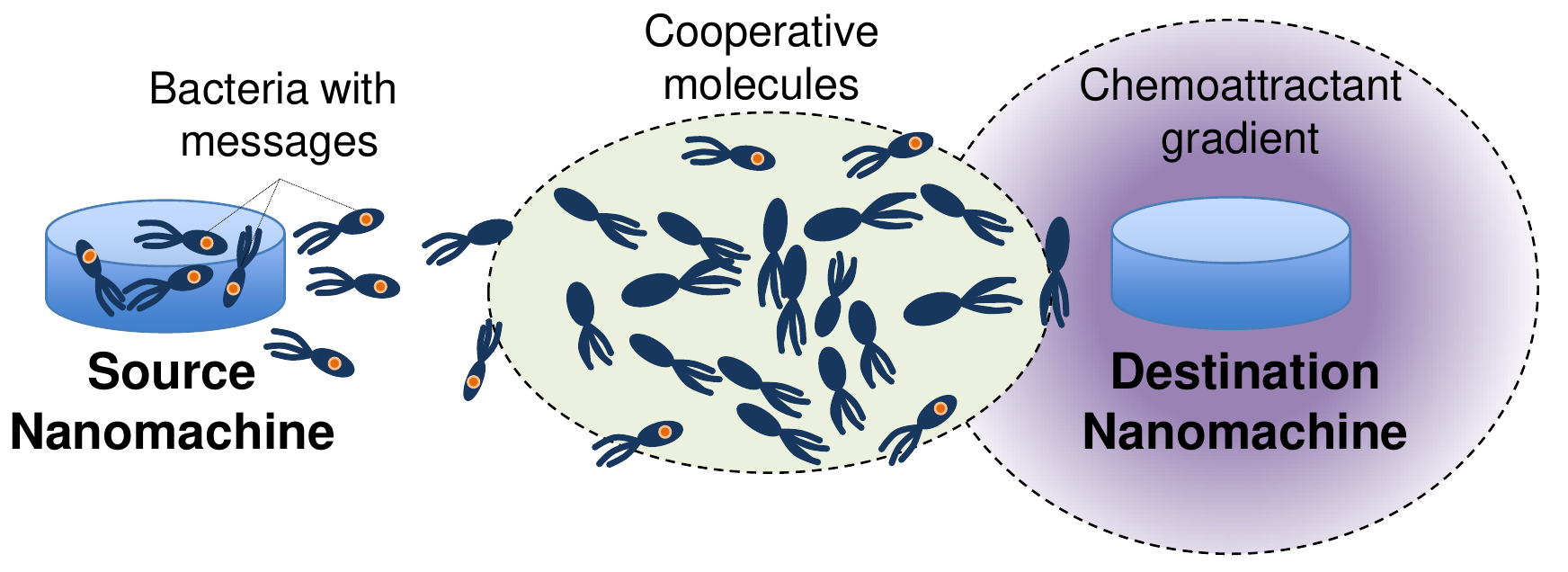}
	\caption{Simulation scenario for the bacterial social behavior.}
	\label{fig:BacSimNets}
\end{figure}

We simulate a network with two nanomachines, which are the source nanomachine and the receiver nanomachine, separated at some distance $l$ as shown in Fig. \ref{fig:BacSimNets}. We consider \textit{E. Coli} bacteria as the information carrier. For realistic modeling purpose, we use similar simulation parameters used in the earlier studies (e.g., \cite{sasi_bbc}, \cite{bact_mobility_diff_adpt}), by mathematical biologists who have developed the models based on the experimental  results. Since the data message (encoded in DNA plasmid) is embedded in the bacteria, each bacterium can be considered as an individual data packet. We utilize bacterial chemotaxis in order to attract the bacteria to swim toward the destination nanomachine. This is achieved by the destination nanomachine releasing the chemoattractant (e.g., nutrient). Bacteria move through a biased random \textit{running} and \textit{tumbling} process and eventually carry the plasmid to the destination. We assume that the source nanomachine transmits in a time division manner, and if the bacterium does not reach the destination within a fixed timeout duration, the information is considered to be lost. We observe the reliability of the network in terms of the successful transmission probability defined by $\eta = \frac{N_d}{N_s}$, where $N_s$ and $N_d$ denote the total number of bacteria released from the source nanomachine and the number of bacteria that reach the destination nanomachine, respectively.

Among the different bacterial social interactions, here we consider the cooperative communication process by means of QS. The cooperative process is established when the bacterium observes increasing chemoattractant density and notifies the others through diffusion of cooperative signaling molecules. The objective is to entice the bacteria carrying the message to bias its directional mobility towards the destination. We assume that in our environment, there is no supporting architecture (e.g., nanotube) between nanomachines and the bacteria are freely swimming in the medium. We model the bacterial mobility \cite{bact_mobility_diff_adpt} as follows:
\begin{equation}
\mathbf{p}_n(t) = \mathbf{p}_n(t-1) + \vartheta \frac{\mathbf{v}_n(t)}{\parallel \mathbf{v}_n(t) \parallel} \chi_n(t) + \mathbf{b}_n(t)
\end{equation}
where $\mathbf{p}_n(t)$ denotes the position of the bacterium $n$ at time $t$ within the timeout duration; $\mathbf{v}_n(t)$ and $\parallel \mathbf{v}_n(t) \parallel$ denote the direction and magnitude of the bacteria movement,  $\vartheta$ is the step size of the bacterium during one time interval, $\mathbf{b}_n(t)$ is an i.i.d. Gaussian random vector representing the tumbling effect and the {\em Brownian motion}. Brownian motion refers to the random collision of molecules in the medium. Due to Brownian motion, even in running mode the direction of the bacterium will change in a random manner. The binary decision variable $\chi_n(t)$ determines whether the bacterium will run or tumble at a time instance $t$. At each time instance, the bacterium decides whether it will run or tumble based on its own ability to make a decision and the information obtained from the environment (e.g., from other bacteria). 
If the sequence of decisions $\chi_n(t)$ for $t = 1, 2, \ldots, T$ eventually leads the bacterium to the destination nanomachine within the timeout duration $T$, the information transmission process is considered to be successful.

Note that, as mentioned in Section \ref{subsec:coop}, bacteria can release cooperative molecules and by sensing the density of the molecules (released by other bacteria), a bacterium biases its mobility accordingly. For the case of cooperative communication, the decision sequence is determined based on both the chemoattractant density and the cooperative molecular signals released by individual bacterium during the QS process. When there is no interaction among the bacteria, the decision sequence is determined based on only the density of chemoattractant observed from the medium.

We consider a steady-state chemoattractant density (e.g., the density of the chemoattractant will not change over time). However, the observed density by the bacterium will vary according to the distance between the current position of the bacterium and the chemoattractant source (e.g., receiver nanomachine). We also assume a stable environment with sufficient nutrients. Therefore, this will lead to minimal non-cooperative behaviors and competition of nutrients among the bacteria, and the bacterial behavior will not change during the communication process.

\subsection{Simulation Results}


\begin{figure*}
\centering
\subfigure[]{\includegraphics[width=2.25in]{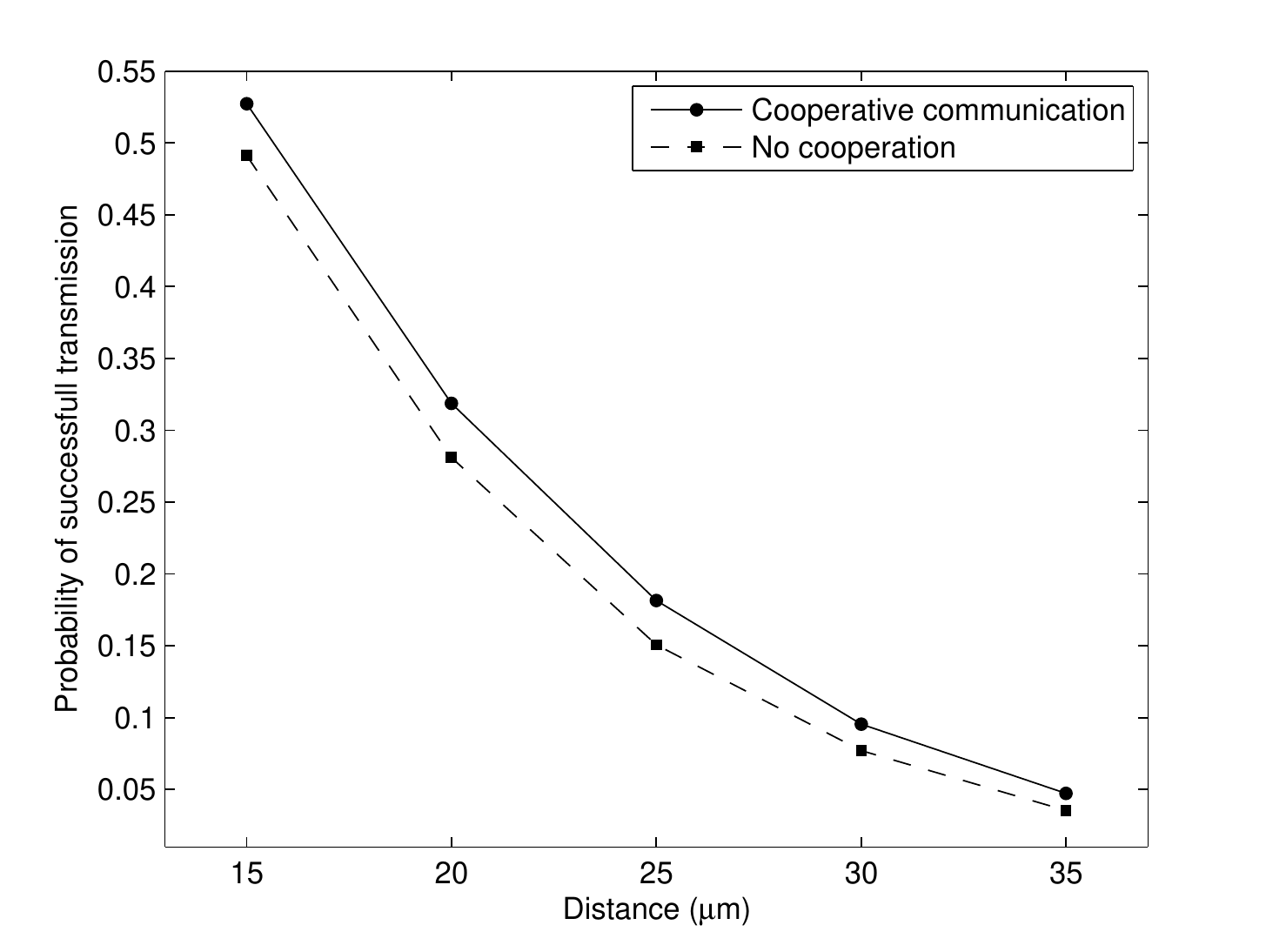}
\label{fig:distance_eff}}
\subfigure[]{\includegraphics[width=2.25in]{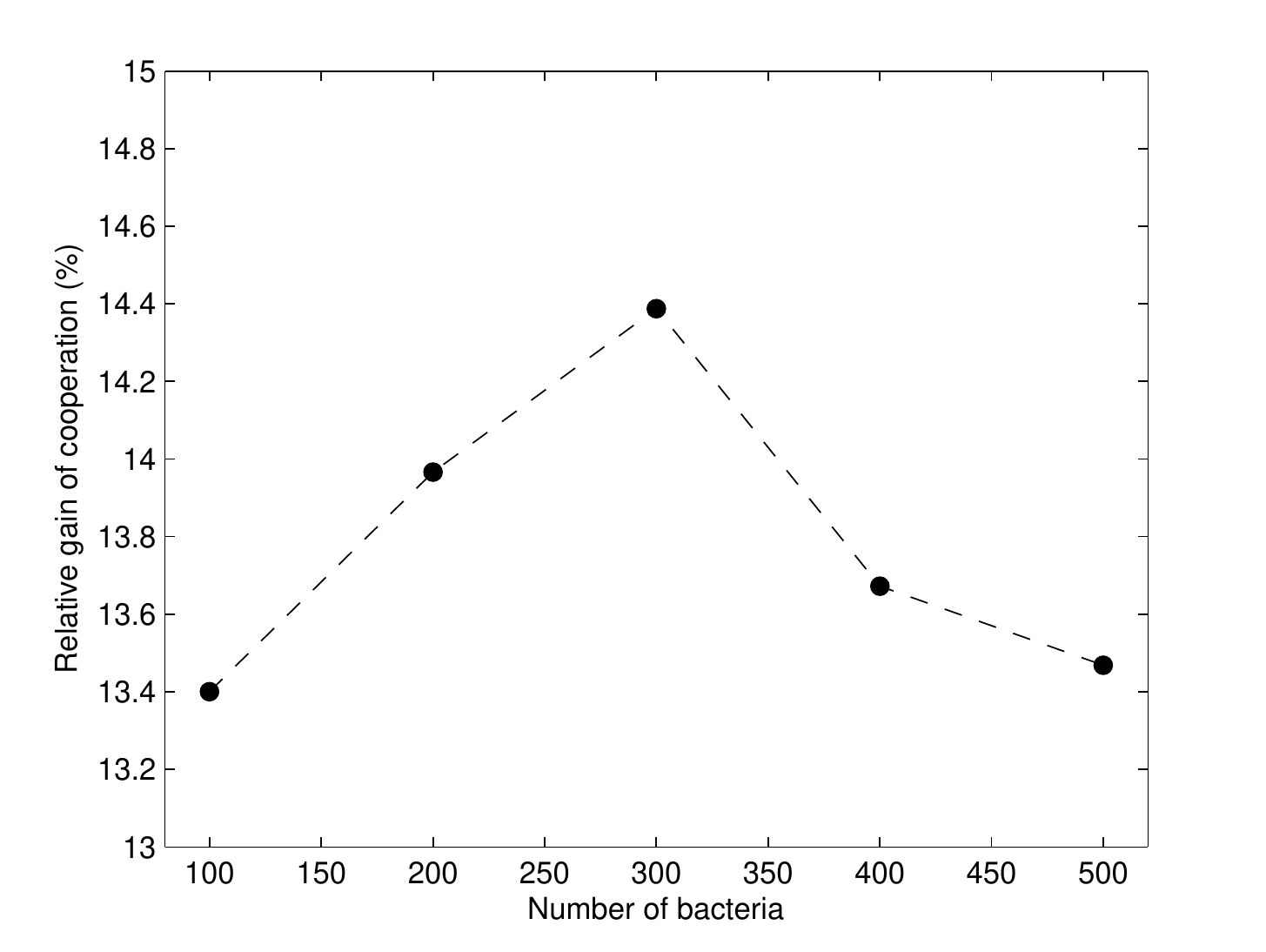}
\label{fig:bacteria_dyn}}
\subfigure[]{\includegraphics[width=2.25in]{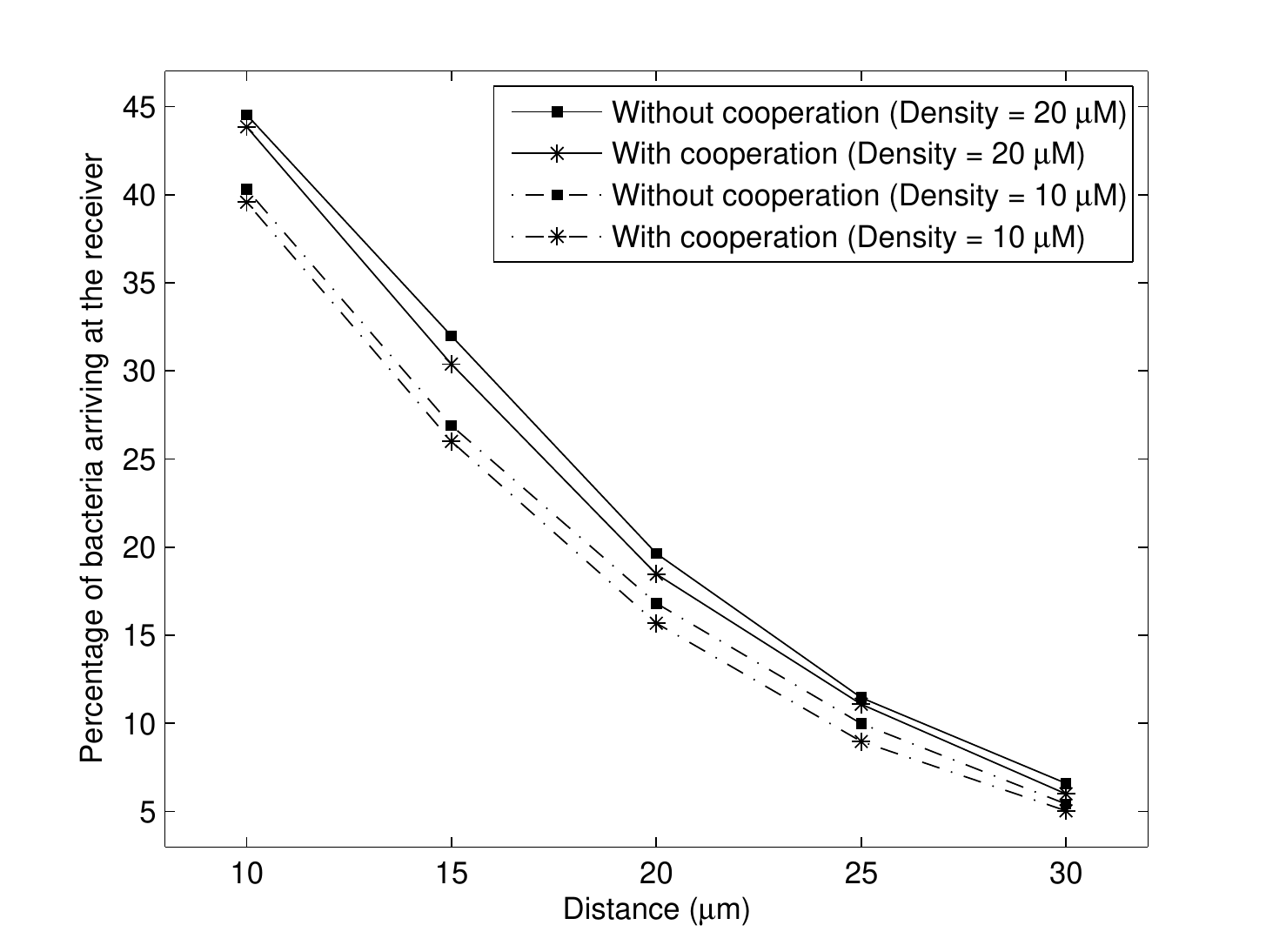}
\label{fig:pop_dyn}}
\caption{Impact of bacterial social behavior on a single-link nanonetwork:  (a) Network reliability with varying distances between the source and destination nanomachines. Total $N_s = 100$ bacteria are released from the source nanomachine and each bacterium carries a single plasmid. The chemoattractant density is assumed to be $10~ \mu \text{M}$. (b) Impact of  the size of the bacterial population on relative gain of cooperation; the distance between the source and destination nanomachines is set to $20~ \mu \text{m}$. (c) Individual dynamics of bacterial cooperation under different density of the  chemoattractant nutrient where $N_s$ is set to $100$. For all the simulations, the timeout duration is set to $1000$ milliseconds. If a bacterium is unable to reach the destination within the timeout duration, the information is considered lost. }
\end{figure*}

\subsubsection{Effect of distance}
In Fig. \ref{fig:distance_eff}, we vary the distance between the source and destination nanomachine. For a fixed timeout duration, when the distance between the nanomachines is high, the bacteria are unable to reach the destination which reduces the probability of successful transmission. Note that, even when there is no cooperation [e.g., dotted curve in Fig. \ref{fig:distance_eff}], a small number of bacteria can still reach the destination using their own sensing abilities (e.g., utilizing the chemotaxis process). Cooperative communication among bacteria helps to attract them toward the chemoattractant gradient. For example, a bacterium obtains additional information about the chemoattractant sources from the other bacteria in the environment and adapts its decision of running and tumbling accordingly. 
However, at larger distances between the source and destination, the effect of cooperation is less prominent due to the fact that the cooperative signaling molecules spread too far and have minimal influence on the bacteria sensing. 

\subsubsection{Effect of changes in the bacterial population}

In Fig. \ref{fig:bacteria_dyn}, we vary the number of bacteria and observe the effect of cooperative signaling molecules on the communication performance in terms of successful transmission probability. We define the relative gain of cooperation as $\Delta_c = \frac{\eta_{cc} - \eta_{nc}}{\eta_{nc}}$, where $\eta_{cc}$ and $\eta_{nc}$ denote the observed network reliability due to cooperative communication and without cooperation, respectively. Note that although increasing the number of cooperating bacteria improves the relative gain, there comes a point when the cooperative behavior leads to a declining gain. For example, when the number of bacteria $N_s = 500$, the relative gain of cooperation is around $13.6\%$, which is 
considerably less than that when the population size $N_s = 300$ (e.g., $14.4\%$). In such a scenario, although increasing the number of bacteria results in an increased number of bacteria at the receiver,  the reliability (in terms of successful transmission probability) does not increase substantially, which leads to a lower gain.

\subsubsection{Individual bacterial behavior}

Fig. \ref{fig:pop_dyn} shows the individual bacterial behavior with different chemoattractant density profile. We consider a situation where a fraction of the population cooperates by producing cooperative signaling molecules that bias the bacteria toward the chemoattractant gradient. However, the rest of the population use only the chemoattractant gradient. Note that, increasing the distance limits the success probability. 
When the nutrient density is reduced to half (i.e., from $20~ \mu \text{M}$ to $10 ~\mu \text{M}$, as represented in the dotted curves), the success probability drops significantly, especially for shorter distances. For shorter distances, the bacteria are close to the chemoattractant source. As a result, the effect of changes in the cooperative signaling molecules is more prominent.

In low-density scenarios, the bacteria are unable to observe the gradient of the chemoattractant nutrient (and hence also fail to signal and cooperate with the other bacteria) which leads to a lower success rate. An interesting observation is that the percentage of bacteria that do not participate in the cooperation, but are able to reach the destination, is higher compared to the percentage of bacteria that cooperate and reach the destination. We can explain this fact as follows: the bacteria that are not part of the cooperative group, can still benefit from the diffused molecules released by the cooperative bacteria. This demonstrates how the non-cooperative bacteria can benefit from the cooperative bacteria.  In addition to their own sensing capability, the bacteria that do not cooperate, also benefit from others' diffused information. As a result, a higher percentage of bacteria will arrive at the destination.
Note that even though certain bacteria diffuse cooperative  signaling molecules to the other bacteria, this does not guarantee that those bacteria will reach the destination.

\subsubsection{Effect of changes in the chemoattractant density}

\begin{figure}[ht]
\centering
\subfigure[]{%
\includegraphics[width=2.25in]{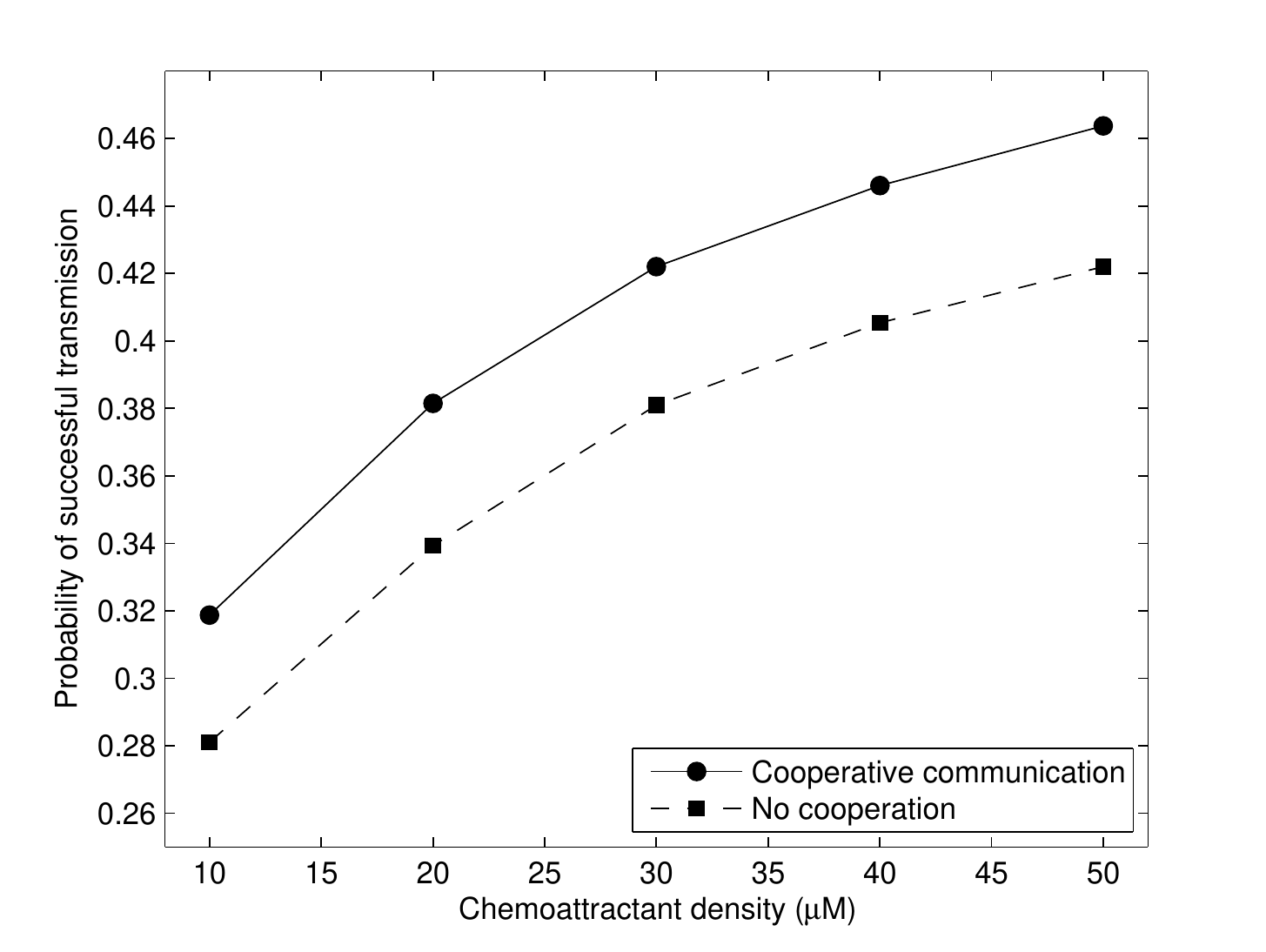}
\label{fig:density_vs_prob}}
\quad
\subfigure[]{%
\includegraphics[width=2.25in]{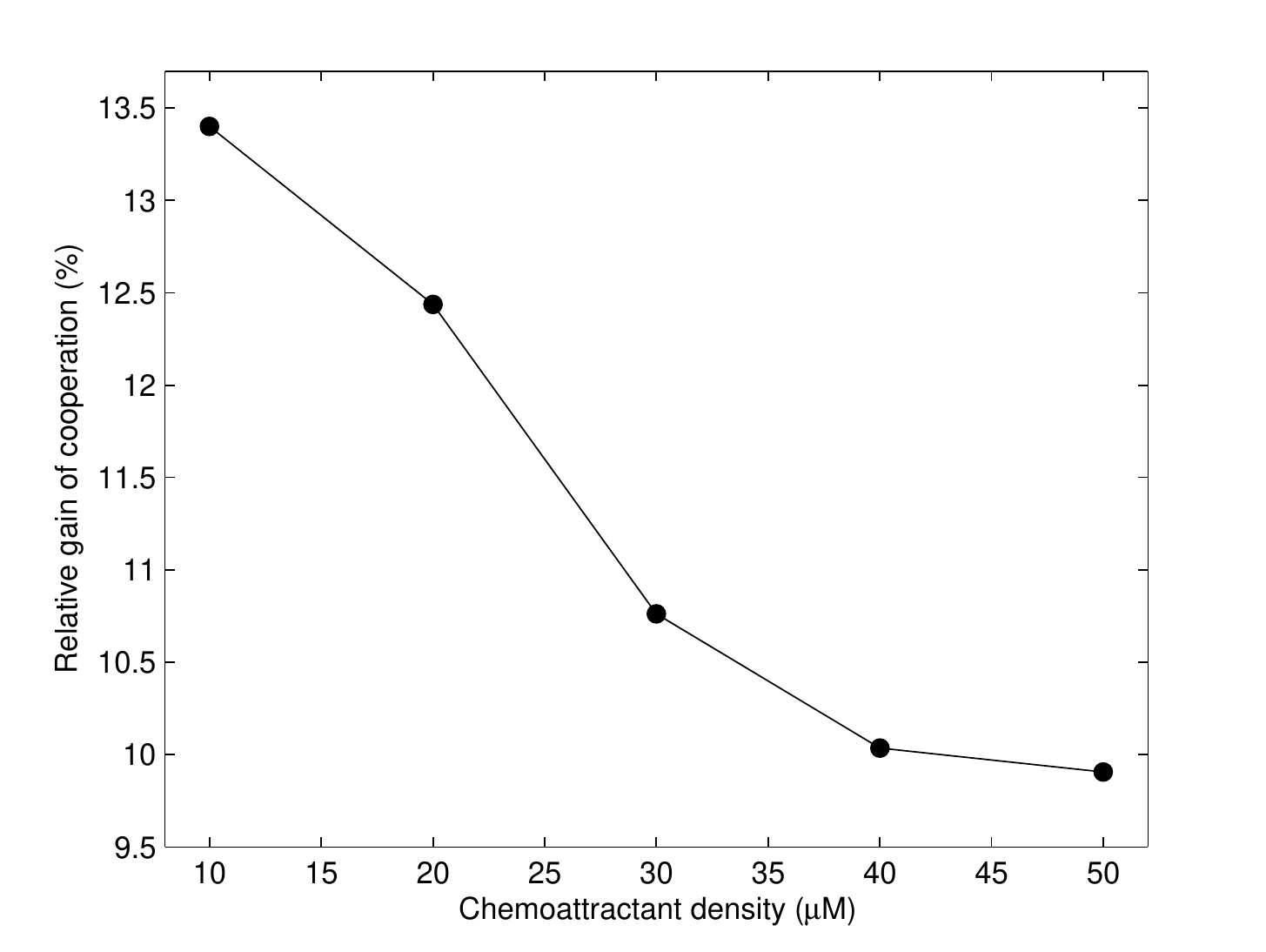}
\label{fig:density_vs_gain}}
\caption{Impact of changes in the chemoattractant density. In (a) we vary the quantity of chemoattractant density and determine the successful transmission probability. In (b) we show the relative gain due to the cooperative signaling molecules with respect to the varying chemoattractant density. Here $N_s = 100$, the timeout duration is set at  $1000$ milliseconds, and the source to destination distance is set to $20~ \mu \text{m}$. }
\label{fig:chemo_den}
\end{figure}

Fig. \ref{fig:chemo_den} shows the communication performance under varying chemoattractant density. In Fig. \ref{fig:density_vs_prob}, as  the density of the chemoattractant increases, this leads to a higher success rate of information transfer. In high-density conditions, the bacteria are able to sense the gradient of the chemoattractant more rapidly which enables them to reach the destination successfully. However, we can still see the benefits of cooperative signaling which helps to bias the directional movement of the bacteria toward the destination. 

The relative gain in terms of the successful transmission probability (due to cooperation) with varying chemoattractant density  is illustrated in Fig. \ref{fig:density_vs_gain}. During the low-density conditions, the effect of cooperative communication is more significant. We can attribute this to the fact that under low chemoattractant density conditions, the bacteria are unable to sense the chemoattractant gradient efficiently, especially when they are far from the chemoattractant source. In such cases, the cooperative signaling molecules aid and compensate for the low chemoattractant density, leading to higher gains. Although cooperative signaling molecules help the bacteria compared to the case when there is no cooperation, 
its  influence on the bacteria is far less compared to a situation with a high density of chemoattractant. 

\section{Outlook on Future Research Directions} \label{sec:fut_dir}

Among numerous research opportunities that will emerge from this new multi-disciplinary research field, we list out a few examples below.

The first research opportunity is the increased research synergy between ICT researchers and molecular biologists, in particular, for the development of wet lab experimental platforms. The \emph{NSF MoNaCo} project\footnote{\url{http://www.ece.gatech.edu/research/labs/bwn/monaco/index.html.}} 
has began developing an experimental platform that brings together communication engineers, microfluidic experts, and molecular biologists. However, the project is only limited to validating bacterial nanonetworks using molecule-based communication. Therefore, future wet lab experimental validations can take on the DNA-based communication in bacterial nanonetworks. By developing experimental platforms for DNA-based communication,  a new collaborative synergy can be established between ICT researchers, experimental bacteriologists, and biotechnologists. The experimental validations can lead to the potential applications that have been described in the introduction, e.g., environmental sensing, biofuel quality monitoring, or new solutions for personalized health-care. 

Another research prospect is to integrate the bacterial nanonetworks with the established solutions found in present nanotechnology research and/or industrial products. A number of research efforts have been dedicated to produce nanoscale components that can be assembled into nanomachines. These nanomachines can perform limited functionalities such as sensing and releasing drug payloads to the diseased cells. Incorporating bacterial delivery process through the nanonetworks can enhance the probability of delivering the elements to the targeted location. 

Lastly, the area of bacterial nanonetworks along with molecular communication can play a major role in the field of synthetic biology. The objective of synthetic biology is the development of artificial creation of biological components and systems that are tailored to perform specific functions. Therefore, using existing knowledge and tools in synthetic biology can help design tailored bacterial nanonetworks that have a certain performance reliability for a specific application.

\section{Conclusion} \label{sec:conclusion}

The use of bacteria as an information carrier has been proposed for molecular communication. Utilizing the bacterial properties such as their ability of carrying plasmids (this could represent the information that has been encoded) and their mobility, could enable information to be transferred between the different nanomachines. Similar to most organisms, bacteria also exhibit social properties, which include both cooperative and non-cooperative behavior. In this article, we have presented an overview of the various communication mechanisms as well as the social properties of bacteria. We have discussed the challenges that arise due to these mechanisms which can affect the information transfer performance in the bacterial nanonetworks. In particular, the challenges due to non-cooperation and opportunities due to cooperation have been  discussed. These opportunities can be exploited in designing nanomachines.  For example, the cooperative and non-cooperative behaviors can be modeled using \textit{game theory} and the bacterial nanonetworks can be engineered to achieve the optimal outcome, for example, by using \textit{mechanism design}.

Simulation results have been presented to evaluate the impact of bacterial cooperative behavior in improving the information transfer performance in a single-link nanonetwork. The results have shown improvement in the communication performance for varying distances between the source and destination nanomachines, as well as situations when the chemoattractant density is varied.

The solutions to the fundamental research challenges in conventional ad hoc networks, such as social-based DTNs, can provide lessons for analyzing communication networks at the nanoscale (e.g., bacterial nanonetworks). The commonality between these two different networks is that the nodes and the organisms, respectively, which carry the information, exhibit social behavior.  A new direction of research to address the research challenges in future social-based molecular nanonetworks can thus be envisaged. 

\section*{Acknowledgment}
\addcontentsline{toc}{section}{Acknowledgment}
This work was supported in part by a Discovery Grant from the Natural Sciences and Engineering Research Council of Canada (NSERC), in part by the Academy of Finland FiDiPro program ``Nanocommunication Networks,'' 2012–2016, and in part by the Academy Research Fellow program (project no. 284531). 

\bibliographystyle{IEEEtran}


\begin{IEEEbiographynophoto}{Monowar Hasan}
(S'13) is currently working toward his M.Sc. degree in the Department of Electrical and Computer Engineering at the University of Manitoba, Winnipeg, Canada.  He has been awarded the University of Manitoba Graduate Fellowship. Monowar received his B.Sc. degree in Computer Science and Engineering from Bangladesh University of Engineering and Technology (BUET), Dhaka, in 2012. His current research interests include Internet of things, wireless network virtualization, and resource allocation in 5G cellular networks. He served as a reviewer for several major IEEE journals and conferences.
\end{IEEEbiographynophoto}

\begin{IEEEbiographynophoto}{Ekram Hossain}
(S'98-M'01-SM'06) is currently a Professor in
the Department of Electrical and Computer Engineering at University of Manitoba, Winnipeg,
Canada. He received his Ph.D. in Electrical Engineering from University of Victoria, Canada,
in 2001. Dr. Hossain's current research interests include design, analysis, and optimization of
wireless/mobile communications networks, cognitive radio systems, and network economics. He
has authored/edited several books in these areas
(\url{http://home.cc.umanitoba.ca/~hossaina}). Dr. Hossain serves as the Editor-in-Chief for the \textit{IEEE Communications Surveys and
Tutorials}, an Editor for \textit{IEEE Wireless Communications}. Also, currently he
serves on the IEEE Press Editorial Board. Previously, he served as the Area
Editor for the \textit{IEEE Transactions on Wireless Communications} in the area
of ``Resource Management and Multiple Access'' from 2009-2011, an Editor
for the \textit{IEEE Transactions on Mobile Computing} from 2007-2012, and an
Editor for the \textit{IEEE Journal on Selected Areas in Communications} - Cognitive
Radio Series from 2011-2014. Dr. Hossain has won several research awards
including the University of Manitoba Merit Award in 2010 and 2014 (for
Research and Scholarly Activities), the 2011 IEEE Communications Society
Fred Ellersick Prize Paper Award, and the IEEE Wireless Communications
and Networking Conference 2012 (WCNC'12) Best Paper Award. He is a
Distinguished Lecturer of the IEEE Communications Society for the term
2012-2015. Dr. Hossain is a registered Professional Engineer in the province
of Manitoba, Canada.
\end{IEEEbiographynophoto}

\begin{IEEEbiographynophoto}{Sasitharan Balasubramaniam}
(SM'14) received his Bachelor (electrical and electronic engineering) and Ph.D. degrees from the University of Queensland in 1998 and 2005, respectively, and the Master’s (computer and communication engineering) degree in 1999 from Queensland University of Technology. He is currently a senior research fellow at the Nano Communication Centre, Department of Electronic and Communication Engineering, Tampere University of Technology (TUT), Finland. Sasitharan was the TPC co-chair for \textit{ACM NANOCOM} 2014 and \textit{IEEE MoNaCom} 2011. He is currently an editor for \textit{IEEE Internet of Things} and Elsevier's \textit{Nano Communication Networks}. His current research interests include bio-inspired communication networks, as well as molecular communication.
\end{IEEEbiographynophoto}

\begin{IEEEbiographynophoto}{Yevgeni Koucheryavy}
 (SM'08) is a Full Professor and Lab Director at the Department of Electronics and Communications Engineering at the Tampere University of Technology (TUT), Finland. He received his Ph.D. degree (2004) from the TUT. Yevgeni is the author of numerous publications in the field of advanced wired and wireless networking and communications. His current research interests include various aspects in heterogeneous wireless communication networks and systems, the Internet of Things and its standardization, and nanocommunications. Yevgeni is an Associate Technical Editor of \textit{IEEE Communications Magazine} and Editor of \textit{IEEE Communications Surveys and Tutorials}.
\end{IEEEbiographynophoto}

\end{document}